\documentclass[12pt,preprint]{aastex}

\tighten
\def\spose#1{\hbox to 0pt{#1\hss}}
\def\ltwig{\mathrel{\spose{\lower 3pt\hbox{$\mathchar"218$}}
     \raise 2.0pt\hbox{$\mathchar"13C$}}}
\def\gtwig{\mathrel{\spose{\lower 3pt\hbox{$\mathchar"218$}}
     \raise 2.0pt\hbox{$\mathchar"13E$}}}

\newcommand{\beq}{\begin{equation}}
\newcommand{\eeq}{\end{equation}}
\newcommand{\beqa}{\begin{eqnarray}}
\newcommand{\eeqa}{\end{eqnarray}}

\newcommand{\simle}{\mbox{$\stackrel{<}{_{\sim}}$}}
\newcommand{\simge}{\mbox{$\stackrel{>}{_{\sim}}$}}

\shorttitle{The prototype colliding-wind pinwheel WR~104}
\shortauthors{ Tuthill et al.}

\begin{document}


\title{The prototype colliding-wind pinwheel WR~104}


\author{Peter~G.~Tuthill}
\affil{School of Physics, University of Sydney, NSW 2006, Australia} 
\email{p.tuthill@physics.usyd.edu.au}

\author{John~D.~Monnier}
\affil{University of Michigan at Ann Arbor, Department of Astronomy, 
500 Church Street, Ann Arbor, MI 48109-1090, USA}
\email{monnier@umich.edu}

\author{Nicholas~Lawrance}
\affil{School of Aerospace, Mechanical and Mechatronic Engineering, 
       University of Sydney, NSW 2006, Australia}
\email{n.lawrance@acfr.usyd.edu.au}

\author{William~C.~Danchi}
\affil{NASA Goddard Space Flight Center, Infrared Astrophysics,
                 Code 685, Greenbelt, MD 20771, USA}
\email{wcd@iri1.gsfc.nasa.gov}

\author{Stan~P.~Owocki}
\affil{Bartol Research Institute, University of Delaware,
       Newark, DE 19716}
\email{owocki@bartol.udel.edu}

\and

\author{Kenneth~G.~Gayley}
\affil{University of Iowa, Iowa City, IA 52245}
\email{ken.gayley@gmail.com}

%
%



\begin{abstract}

Results from the most extensive study of the time-evolving dust structure
around the prototype ``Pinwheel'' nebula WR~104 are presented.
Encompassing 11 epochs in three near-infrared filter bandpasses, a homogeneous 
imaging data set spanning more than 6 years (or 10 orbits) is presented.
Data were obtained from the highly successful Keck Aperture Masking Experiment,
which can recover high fidelity images at extremely high angular resolutions, 
revealing the geometry of the plume with unprecedented precision. 
Inferred properties for the (unresolved) underlying binary and wind system are 
orbital period 241.5$\pm$0.5\,days and angular outflow velocity of 
0.28$\pm$0.02\,mas/day.
An optically thin cavity of angular size $13.3\pm1.4$\,mas was found to lie
between the central binary and the onset of the spiral dust plume.
Rotational motion of the wind system induced by the binary orbit is found to have 
important ramifications: entanglement of the winds results in strong shock 
activity far downstream from the nose of the bowshock.
The far greater fraction of the winds participating in the collision may play a
key role in gas compression and the nucleation of dust at large radii from the 
central binary and shock stagnation point.
Investigation of the effects of radiative braking pointed towards significant 
modifications of the simple hydrostatic colliding wind geometry, extending the 
relevance of this phenomena to wider binary systems than previously considered.
Limits placed on the maximum allowed orbital eccentricity of $e$~\simle~0.06 
argue strongly for a prehistory of tidal circularization in this system.
Finally we discuss the implications of Earth's polar ($i$\simle$16^\circ$) vantage
point onto a system likely to host supernova explosions at future epochs.


\end{abstract}


\section{Introduction}

The first well-resolved images of the dusty, IR-luminous Wolf-Rayet
star WR~104 revealed an elegant plume stretching hundreds of AU from the 
bright core and following a trajectory closely matched to an Archimedeian 
spiral \citep{nature99}.
The origin of this geometry is simple and highly intuitive: material is 
embedded within a uniformly-expanding spherical wind driven from the hot
stars at the heart of the system, yet the orbital motion of a central
binary causes a rotating wake embedded within the flow.
These images gave immediate confirmation of the colliding-wind binary nature 
of dust production in WR~104, and at the same time presented astronomy with 
a new type of object for study: a ``Pinwheel'' nebula.

Although we found no specific predictions in the literature for such a 
spiral dust plume in dusty Wolf-Rayets such as WR~104, the same combination of 
orbital and outflow motions led \citet{Kuiper41} to predict exactly this geometry 
for the circumstellar environment around the eclipsing binary $\beta$~Lyr.
In fact use of the term ``Pinwheel'' to describe such a structure can be found 
as early as 1950 \citep{Struve50}, and theoretical study of colliding-wind
binary systems has since developed an extensive literature 
\citep[e.g.][and refs therein]{SB92,WF00}.
The basic picture of colliding-winds mediating dust production was formulated 
to explain the episodic dust-producer WR~140, which became something of a
`Rosetta Stone' system in important papers establishing the model 
\citep{Williams90,Usov91}.
It is also interesting, although perhaps tangential to the work of this paper, 
to note that Archimedian spirals may arise in quite distinct mass-loss environments
such as the binary reflex-motion induced structure recently seen in AFGL~3068
\citep{MH06}.

Observational confirmation of the pinwheel nebula around WR~104 has not yet been 
attained by adaptive optics imaging due to the requirement for a stable, well 
characterized point-spread function enabling recovery of structure at the diffraction 
limit \citep{Jay04}.
However, a number of other techniques have been used to confirm the presence of the
spiral plume including high spatial resolution lunar occultations \citep{MS02}, 
variations in the visible lightcurve of 241\,days in accord with the orbital period 
derived from interferometry \citep{Kato02}, together with extensive follow-on 
observations with aperture masking interferometry \citep{wrconf02,wrconf03,Monnier07}.
Reports of binarity in the WR~104 system \citep{Crowther97} appeared shortly before 
the original discovery of the spiral, with the spectral classification for the object 
given in \citet{7cat} as WC9d~+~B0.5V~(+VB), although given the uncertainties we refer 
to the second component spectrum (actually brighter by a factor of $\sim$2) as OB.
Detailed radiative transfer modelling, confirming the appearance and basic observational
properties of the pinwheel, has been performed by \citet{Harries04}.

The discovery of similar structures in an expanding list of host systems
has served to motivate further interest in these objects.
The second such system to be found, WR~98a \citep{Monnier99}, showed a similar 
morphology implying approximately matched basic properties (in terms of orbital period 
and windspeed) to the prototype WR~104.
However, a more diverse population is also coming to light including WR~112 
\citep{Marchenko02} which appears to have a significantly longer 25\,yr binary 
period, and the high orbital eccentricity 8\,yr system WR~140 whose periodic 
periastron passages have been shown to throw off a remarkable series of dusty arcs 
\citep{Monnier02b,MM06}.
Recently, a dramatic association of at least two and probably five such spiral
systems was discovered near the Galactic Center at the heart of the massive
Quintuplet Cluster \citep{Quint06}.

The status of Wolf-Rayets as candidate progenitors to type~Ib/c  
supernovae \citep[reviews can be found in][]{Bethe90,VLV98,WHW02} has generated
considerable interest in their mass loss histories.
In particular, structures within the circumstellar environment, into which the
supernova blast wave expands, have the capacity to interact with the ejecta modifying
the lightcurve and observed properties of the explosion. 
Indeed, \citet{Ryder04} observed a residual fluctuating or rippled signal, obtained
when subtracting a spherical best-fit model from the observed radio lightcurve of
supernova 2001ig.
The most likely interpretation for these modulations, they argue, arises as the
supernova shock encounters periodic density enhancements embedded in the 
circumstellar environment by a colliding-wind binary pinwheel system.
With the rapidly solidifying links between supernovae and gamma-ray bursts
\citep{WB06}, the mechanisms by which such mass-loss signatures may become
encoded within the lightcurves of the most energetic explosions in the
cosmos are being explored \citep{MR05}.

In this paper, we present the results of the most extensive diffraction-limited
imaging study yet performed on the prototype pinwheel system, WR~104. 
Observations encompass three different near-infrared filter bandpasses and eleven 
distinct epochs covering more than six years (almost ten orbital periods) of the 
system.
With the structure of the time-evolving plume revealed in unprecedented detail, 
we are able to address key outstanding questions regarding the way in which the
dust formation is mediated by the colliding winds of the binary.
Multi-wavelength observations are helpful in teasing apart the complex
three-dimensional geometrical, thermal and illumination effects which go into 
creating the phenomena we see as a pinwheel nebula.

\section{Observations}

Monitoring of a number of dust-forming Wolf-Rayet stars was performed as
part of our aperture-masking interferometry program begun at the Keck~1 
telescope in 1996.
This has produced astrophysical results spanning a range of different areas
within contemporary stellar physics including studies of young stars
\citep{mwc349,Tuthill02}; evolved pulsating stars and giants 
\citep{Tuthill00,Monnier04}; and dusty mass-loss shrouds in proto-planetary
nebulae and transition objects \citep{Monnier00,redrect02}.
A full description of the experiment including a discussion of the conceptual
principles and signal-to-noise considerations underlying masking interferometry
is given in \citet{keckmask} while further discussion of systematics and seeing
induced errors can be found in \citet{Monnier04}.

Observations presented here all used a partially-redundant annular mask geometry 
first suggested by \citet{HB92} which was found to deliver robust imaging for
targets as faint as $m_K\sim6$\,mag with complete Fourier coverage out to the
8\,m maximum baseline passed by the mask.
The detector used for this work was the NIRC camera with image magnifying 
front-optics modifying the plate scale so as to sample the diffraction-limited
beam \citep{nirc96}.
A relatively rapid-exposure ($T_{int} = 0.14$\,sec) and high data volume (data
cubes containing 100 frames) speckle observing mode was utilized, although this
was not as well developed as it could have been and suffered from an inherent low
duty cycle of at best 20\% set by the camera electronics. 

Observations of science targets were interleaved with those of nearby unresolved 
point-spread function (PSF) reference stars, and the extraction of calibrated
visibility and closure-phase data followed established practices within speckle
or masking interferometry \citep{keckmask}.
All $V^2$ and closure phase data products were stored in the FITS-based 
Optical Interferometry data exchange format \citep[OI-FITS][]{oifits05}, 
and are available on request from the authors.
Images were recovered using a number of different numerical methods 
including the CLEAN algorithm \citep{clean} and with several maximum
entropy \citep{mem} based methods \citep{devinder,macim}.
With some variation in levels of fidelity and resolution obtained, all methods 
produced similar structures to those presented in the following sections, and
hereafter we show exclusively the results of the VLBMEM package \citep{devinder}.

A journal of masking observations of WR~104 is given in Table~\ref{obstable},
while information about the three interference filters within the infrared
H, K and L bands utilized in the study is given in Table~\ref{filtable}.
A number of different PSF reference stars were observed, all of which are relatively
nearby (typically within about $5^\circ$ on the sky) and tailored to give roughly
similar received counts under various observing configurations. 
The primary calibrator stars utilized were SAO~186681, SAO~186841, 14~Sgr and 
HD~165813 although occasionally additional stars were used for confirmatory 
checks.

In addition to their scientific merit, it is worth noting that the images presented
in later sections are a strong testament to the power of aperture masking 
interferometry in the recovery of high fidelity images.
In particular, the pinwheel around WR~104 resolved here possesses strongly 
asymmetric structure at intermediate dynamic range to the bright core.
Deconvolution to obtain at least the full diffraction limit is essential (and 
ideally, some super-resolution beyond it), as the structure
recovered is formally only a few resolution elements across.  
Seeing conditions over the twelve observing nights found in Table~\ref{obstable}
varied over a wide range from extremely poor to very good.
Robust, high-fidelity images were recovered for all epochs and in all filters observed.

This is in contrast to the failure to recover significant reliable 
structure from the one attempt (we are aware of) to image this target with 
adaptive optics \citep{Jay04}.
The presence of asymmetric and time-varying extended structure at high resolutions 
which is nevertheless completely predictable in morphology makes WR~104 almost
ideally suited as a ``test pattern'' target against which imaging experiments may 
be benchmarked, and we look forward to confirmatory results from new
generations of AO instruments.

\subsection{K-band images}

\begin{figure}
\centerline{\includegraphics[angle=180,scale=0.8]{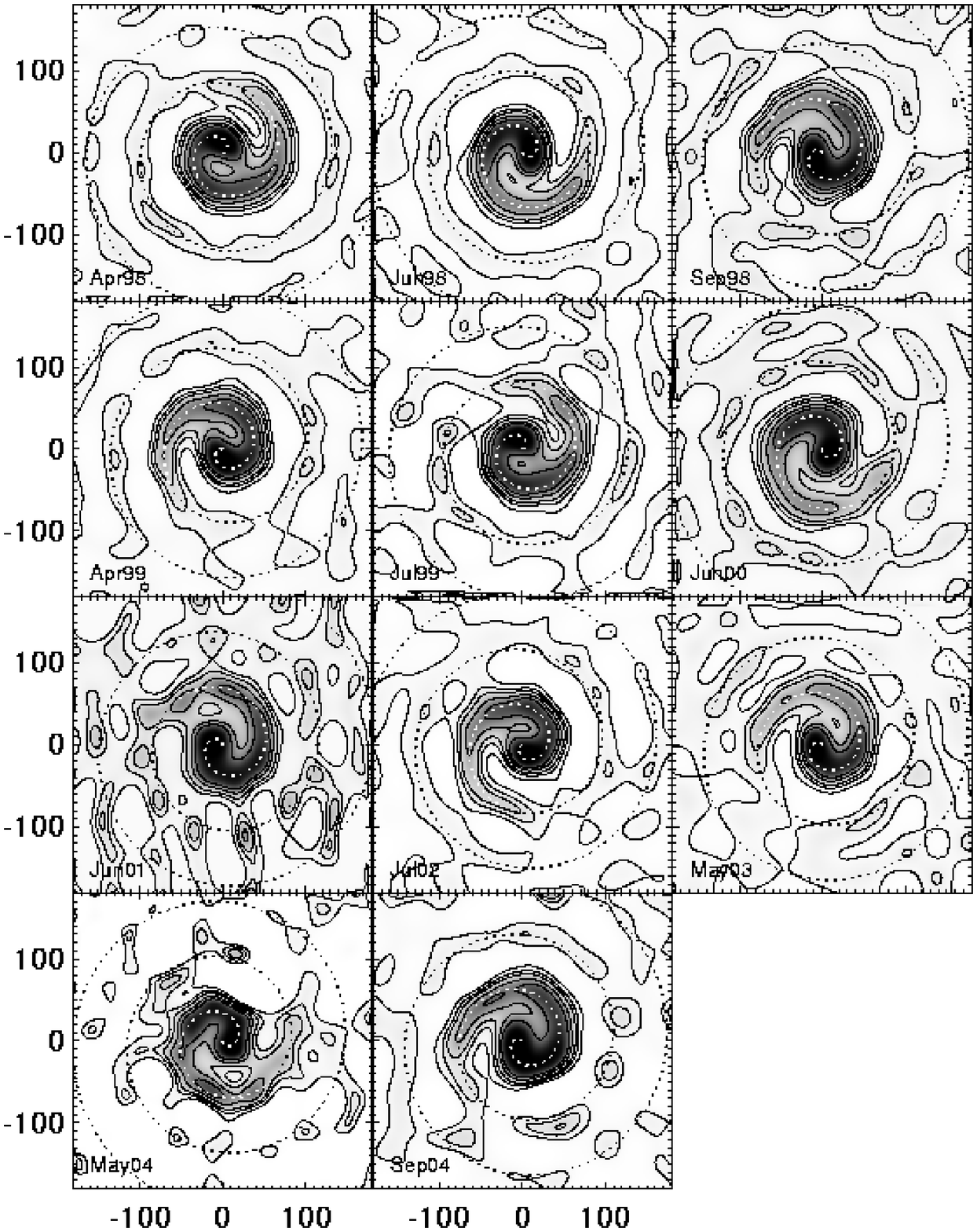}}
\caption{\label{kmaps}
Images recovered from data in the CH4 filter over 11 observing epochs.
Contour levels are .4, 1, 2, 5, 10, 50\,\% of the peak. North (Dec.) 
is up and east (R.A.) to the left with the image scale labelled in 
millseconds of arc. 
Images have been centered on the mathematical origin of the 
best-fit Archimedian spiral model (dashed line: see text for details).
}
\end{figure}

Images recovered from data taken at 11 separate epochs in the CH4 filter
are presented in Figure~\ref{kmaps}.
Although there are 12 observing dates given in Table~\ref{obstable}, the
first two of 1998 Apr~14 and Apr~15, being separated by only one day, did
not give significantly different images and these data have been averaged together
for the remainder of this paper.
Experience has taught that images recovered from filters within the infrared
K-band usually deliver the highest fidelity. 
The PAHCS filter at longer wavelengths has significantly lower angular resolution
with a corresponding loss of detail.
On the other hand, the H filter at shorter wavelengths suffers from more severe
seeing-induced calibration noise, and furthermore the target star WR~104 is
also two magnitudes fainter here than at K (due to the rising IR excess).
We therefore begin our discussion of the image morphology with images recovered from
the CH4 filter in the K-band, and we later contrast the results from the other filters 
against this reference.

The images of Figure~\ref{kmaps} demonstrate the recovery of extended structures 
around the bright core with typical image dynamical ranges in excess of 100:1, 
although higher noise levels due to poor seeing did limit fidelity on occasion
(e.g. Jun01 and May04 epochs).
The predominant feature at all epochs is a curved plume originating in a bright 
compact core, and wrapping around a full turn at which point there is a rapid 
fading in surface brightness to about 1\% of the image peak.
Beyond the bright first coil, at angular radii $\simge$ 100\,mas, there can 
often be seen further windings of a continuing spiral, or in some cases substantial
segments of arcs, at a significance at or somewhat above (e.g. Apr98, Jun98)
the noise level in each individual image.

The overplotted dashed line on each image in Figure~\ref{kmaps} gives the best-fit 
Archimedian spiral model, which is discussed in greater detail in the following 
section.
Each image has been registered so that the mathematical center of the 
best-fit spiral (and not the brightest pixel) defines the coordinate origin.

\subsection{Archimedian spiral model}

An automated procedure was used to extract the coordinates of points lying 
along the crest of the bright spiral ridges from the images of Figure~\ref{kmaps}.
A model could then be fit to all 11 epochs of data simultaneously.
This model consisted of a uniformly-rotating Archimedian spiral, with 
free parameters: $P$ the rotational period, $W$ the angular expansion (wind) speed,
$i$ the inclination to the line-of-sight, $\phi$ the position angle of $i$, and
$\theta_0$ the position angle of the model at the first epoch (Apr98). 
There were two additional translational degrees of freedom associated with the
registration of the model to the image at each epoch.

With nearly 10 complete rotations of the structure covered by our data series, 
it was possible to obtain significantly more precise estimates for the model 
parameters than previously published work.
In particular, best-fit models constrained $P$ to be 241.5$\pm$0.5\,days, a value
in very good agreement with the previous best interferometric value 
\citep[243.5$\pm$3\,d;][]{wrconf02} and with the reported photometric
period \citep[241\,d;][]{Kato02}.
The windspeed $W$ was found to be 0.28$\pm$0.02\,mas/day (or 102$\pm$7\,mas/yr),
also in accord with prior findings \citep{wrconf02}.
Our best fit for $i = 12^\circ$, although the 1-sigma uncertainty covers
a range from $0^\circ$~--~$16^\circ$.
Precise constraint of the inclination was hampered by the nearly face-on 
viewing angle to the system: $12^\circ$ only results in a 2\% distortion 
of the aspect ratio.
Sky orientation parameters were found to be $\phi = 84^\circ\pm15^\circ$
and $\theta_0 = 269^\circ\pm1^\circ$.

Despite the fact that the model was only fit to ridge points extracted from 
the bright inner winding of the spiral, it can be seen from Figure~\ref{kmaps} 
that it also provides a relatively good representation of the structure in
the fainter outer regions.
This implies that the second winding of the spiral follows closely the
expected trajectory from simple expansion, and that the plume remains
relatively well defined and compact as it evolves.

A key finding of the modelling study was that the mathematical origin of the
best-fit spiral was always systematically offset from the center of the 
bright core of flux in the images in a highly consistent fashion,
as can be seen from Figure~\ref{kmaps}.
This ``standoff distance'' between the spiral center and brightest pixel 
was found to be $13.3\pm1.4$\,mas which amounted to a rotational displacement 
downstream along the locus of the spiral of  $85^\circ\pm14^\circ$.
Implications for the physical model of the dusty colliding-wind binary from
the modelling and imaging presented here are discussed further in 
later sections.

\subsection{H and L-band images}
\label{HLimages}

A series of 11 images recovered from H-filter data are presented in Figure~\ref{hmaps}.
Although image fidelity is somewhat lower than for the CH4 filter, the inner bright
spiral structure is well displayed at all epochs. 
Similarly, Figure~\ref{lmaps} shows images recovered from the PAHCS filter from the 
short-wavelength side of L-band. 
Data were only collected in this band for the first six epochs of our study (see
Table~\ref{obstable}), and the resolution obtained is lower than the other filters
due to the longer wavelength.
The spiral structure of the inner coil is again readily apparent.

Archimedian spiral models were fitted to all H and PAHCS images following the procedure
described above.
Model parameters such as rotational period and winding angle were found to be in 
accord with the results from the CH4 filter, however the lower quality of the image
data meant that the fits did not significantly advance our knowledge of these 
properties.

A more profitable strategy for interpretation of the H and PAHCS images was to
fix the spiral model parameters $P$, $W$, $i$ and $\phi$ to their values obtained
above, and to examine the new data for evidence of systematic changes from this 
baseline model.
One relatively subtle change found was in the mean rotational orientation of the 
spiral (model parameter $\theta_0$).
It was found that the mean rotational displacement of models fitting the H images
were retarded by $6^\circ$ (counterclockwise) while the PAHCS images were found
to be advanced by $4^\circ$ (clockwise) compared to the CH4 reference model.
This effect was seen in two independent data-fitting approaches, the first used
a similar strategy to that above in fitting to points along ridge crests, while 
the second entailed the generation of an average CH4 image which was then 
rotated and fit to each trial H/PAHCS image. 
The latter procedure was found to be more robust as it was less sensitive to 
resolution effects which might blur out the exact location of the ridge crest.

A strong note of caution needs to be emphasized here because these relatively small
changes in rotation are obtained from comparison between images with differing 
angular resolutions.
In order to investigate the effect of the decline in image resolution going to
longer wavelengths H--CH4--PAHCS, simulations were performed in which images were 
artificially smoothed, and models fit to determine if there were systematic changes 
in the recovered parameters.
Indeed, we found clear and systematic changes of best fit rotation angle, measured 
to be about $5^\circ$ in a {\em counterclockwise} sense when the simulated 
resolution was degraded to imitate the transitions H--CH4 and CH4--PAHCS.
This makes intuitive sense: lowering the resolution tends to blur the center
together, and the projecting arm of the spiral {\em appears} to rotate as more of it
is lost to the expanding core.
However what is interesting is that this resolution-induced rotational bias 
is in the opposite sense to the observed {\em clockwise} rotations of $6^\circ$  
H--CH4 and $4^\circ$ CH4--PAHCS.
It is therefore likely that the true magnitude of this rotation with observing
wavelength is underestimated here by a factor of two due to the bias counteracting
the signal.
However, the fitting of any models to images at the resolution limit is challenging,
and spirals perhaps even more so as due to their self-similar nature.
The confirmation and unveiling of detailed substructures within the WR~104 plume
must therefore await more powerful telescopes or imaging arrays with still higher 
angular resolution.

\subsection{Lightcurves}

Although our imaging experiment was not specifically designed to yield
flux measurements, it was possible to extract photometry within our three
near-infrared filters to within $\sim 0.1 - 0.2$\,mag, although for some
nights there were light clouds which compromised the measurement accuracy.
These measurements are presented in Figure~\ref{lightcurves}, and show the
infrared fluxes to be fairly constant in the H and CH4 filters (to within a 
few tenths of a magnitude). 
A larger variation is seen in the PAHCS filter, although it lacks any clear
sinusoidal signature with phase and may arise from higher noise levels due
to atmospheric contamination (this filter lies near the edge of the infrared 
L band window).


\section{Basic Geometry and Physical Properties}
\label{basicprop}

In order to illustrate the physical properties of the WR~104 pinwheel
system in the discussions that follow, a cartoon of the basic geometry 
is given in Figure~\ref{cartoon}.
A pole-on view onto a $20^\circ$ opening half-angle spiral shock is depicted, 
together with key elements of the system such as the central binary star,
the optically thin inner region of the plume and the dusty first and
second coils.

If we are able to measure the physical speed of the dust plume, then
this may be combined with the apparent angular velocity to yield the
distance to the system.
Following \citet{nature99}, if we identify the terminal WR windspeed 
obtained spectroscopically as $V_\infty = 1\,220$\,km/s \citep{HS92} then we
may combine this with our proper motion of 0.28$\pm$0.02\,mas/day to yield a 
distance to the system of 2.6$\pm$0.7\,kpc.
By far the dominant error is from the $\sim$25\% uncertainty in the
windspeed found by comparing the results of different line profile studies
\citep{RN95} (velocities as high as 1\,600\,km/s have been reported for
WR~104 \citep{TCM86}).

The distance derived in this fashion is somewhat larger than the 1.6\,kpc 
estimate from possible membership of Sgr~OB1 \citep{LS84}; a discrepancy 
which motivated \citet[][hereafter HMSK]{Harries04} to question the validity 
of this method.
These authors preferred the smaller distance scale on the basis that the 
WR wind may be subject to radiative braking at the OB~star \citep{GOC97}, 
and thus the terminal wind velocity may not be representative of the 
bulk motion of the observed dust plume.
It is instructive to pursue this controversy further, for it illuminates
some of the unique and interesting properties of these Pinwheel systems.

Let us consider a very simple picture in which we distinguish only 3 populations
of gas in the colliding wind system: the free WR wind, the free OB wind, 
and the material around the shock surface which will contain both, but with
modified density and velocity.
We will furthermore, for the present, restrict our attention to the 
hydrodynamic treatment of the inner shock with opening angle given
by HMSK.
This ignores possible modifications to the shock geometry due to radiative 
braking, and we devote the Section~\ref{radbrak} to a discussion of these effects.

For an ordinary (non-rotating) bowshock system, these populations of gas
interact strongly in the heart of the system but at large distances are
able to flow more-or-less unimpeded for they occupy distinct regions of space.
As the dust is known to form around the colliding-wind interface, the arguments 
of HMSK would be correct in suggesting a lower velocity for the  
material in the dust plume from the various braking mechanisms (radiative, shocks).

However, this picture breaks down for the Pinwheel systems when we introduce
the rotation of the plume driven by the binary orbit. 
Recent three-dimensional hydrodynamical simulations of colliding-winds in 
binary systems which include the coriolis effect have shown strong departures
from the simple non-rotating case \citep{LSG07}.
Asymmetric shock strengths on the leading and trailing edges of the shock spiral
were demonstrated.
Although these simulations were mostly concerned with the inner regions of
colliding wind systems, there will also be important consequences at large
radii which need to be considered for the case of WR~104's dust plume.

The resultant geometry will be much more complicated because the OB wind will
become thoroughly wrapped and entangled within the spherically-expanding WR 
wind. 
The interpenetration of the two winds leads to a very different outcome 
than for the non-rotating case.
It is no longer possible for the distinct populations of gas
to travel at different speeds without interfering, and in the orbital
plane containing the shock,  at large radii all 
populations of gas must collide and eventually come to a common velocity.
The two key questions we need to answer to understand the form of the 
spiral plume are (1) what is this final velocity $V_F$, and (2) how long 
will it take until the material in the plume is moving at $V_F$?

We may answer (1) with a quantitative discussion of the wind kinetic energy.
Following HMSK, for the WR component we adopt a mass-loss
rate ${\dot M} = 3 \times 10^{-5} \,M_\odot$/yr, and a windspeed
$V_{\infty,WR} = 1\,220$\,km/s, while for the OB star we have 
${\dot M} = 6 \times 10^{-8} \, M_\odot$/yr and $V_{\infty,OB} \approx 2\,000$\,km/s.
The opening angle for the shock can now be calculated from the formula given
in \citet{EU93} to be $\theta = 17.4^\circ$ half-angle (rounded to $20^\circ$ half
or $40^\circ$ full opening angle in HMSK).
Note that this explicitly ignores any widening of the opening angle due
to radiative braking (see Section~\ref{radbrak}), but we proceed with this
value for the present to ensure consistency with HMSK.
At large radii, the solid-angle subtended by the shock cone will be 
$\Omega = 2\pi(1-cos\theta) = 0.03$\,sr, so that in the non-rotating case,
about 3\% of the WR wind and 97\% of the OB wind will participate in the
collision.
We define $R_{bowshock}$ to be the volume ratio of the WR wind involved in the 
initial bowshock.

In the rotating case, however, a far larger fraction of the WR wind will 
eventually collide due to the entanglement of the winds. 
Winds along the two polar axes, with solid angle 
$\Omega_p = 2\pi(1-cos(\pi/2 - \theta)) = 4.1$\,sr from the WR can propagate 
unimpeded.
However all gas within an equatorial angular band within $\pm \theta$ from
the plane of the orbit must eventually collide with the plume -- for our
geometry of $\theta = 20^\circ$ this corresponds to 34\% of the total WR wind.
Of this 34\%, it is crucial to point out that only $R_{bowshock}$~=~3\% will undergo 
a `prompt' collision at or just downstream of the shock stagnation point, the remaining
31\% will only collide at larger radii as the spiral windings gradually entangle the 
winds.

If we assume, as seems reasonable for a relatively wide colliding-wind binary, that 
the shock is mostly adiabatic with little radiative energy loss. 
The kinetic energy flux of the asymptotic flow for the entangled winds can be found: 

\beq
\label{ke_balance}
\frac{1}{2} {\dot M}_F V_F^2 = 
\frac{1}{2} R_{WR} {\dot M}_{WR} V_{\infty,WR}^2 + \frac{1}{2} R_{OB} {\dot M}_{OB} V_{\infty,OB}^2
\eeq

Here, $R_{WR}$ and $R_{OB}$ are the fractional volumes of the WR and OB winds that are
involved in the collision; from the geometrical arguments above we take 
$R_{WR} = 0.34$ and $R_{OB} = 1.0$. 
Substituting values into this equation we calculate that the fast (but tenuous) OB wind will 
modify the final $V_F$ in the equatorial plane from 1\,220\,km/s to 1\,226\,km/s: an 
insignificant change. 

The model may be further refined by adding a third population of radiatively braked
material, as suggested by HMSK.
Let us suppose (as an upper limit) that {\em all} WR wind impinging upon the 
$\theta = 20^\circ$ conical shock (3\% of the total) is so braked. 
A preliminary estimate of 800\,km/s for $V_P$ the streaming velocity of material in 
the shock cone has been obtained by fitting to the C~III 5696\,\AA\ line profile 
(G. Hill et al., 2007 in preparation).
If we add this third population of gas to Equation~\ref{ke_balance}, we 
obtain $V_F = $1171\,km/s: a decrement of only 4\% on $V_{\infty,WR}$.
We therefore answer part (1) of our question above that $V_F = V_{\infty,WR}$ with
only very minor adjustment, confirming that the momentum-dominant WR wind will 
overwhelm other wind populations.

The second part of the question, finding how long will it take for the WR wind 
to predominate, has a bearing on the shape of the inner spiral: it will be
distorted away from Archimedian if the matter is accelerating.
A rough estimate can be obtained from a simple ``mass loading'' argument applied
in the equatorial plane of the system.
Let us suppose that in the wake immediately behind the OB star, a mass
$R_{bowshock} \times {\dot M}_{WR}$ of matter in the plume has a velocity 
$V_P \ll V_{\infty,WR}$. 
If material were to continue on this trajectory, after time $t$ it would be
overtaken by a mass of the faster wind equal to
\beq
m = (1 - \frac{V_P}{V_{\infty,WR}}) \frac{t}{P} \, R_{WR} \, {\dot M}_{WR}
\eeq
where P is the orbital period. 
We may solve this equation for $t / P$ to give the fraction of an orbital period 
in which a mass of material has collided with the plume from the free WR wind which
is equal to that already present from the initial bowshock. 
Using values from above, we find that 0.1 of an orbital period is sufficient to double
the piled-up mass at the inner wall (see Figure~\ref{cartoon}), and by the time matter
has reached the dust formation zone at a quarter turn, it will be strongly
($\sim$70\%) momentum-dominated by the $V_{\infty,WR}$ wind.

This analysis confirms the general picture that the spiral plume in this system
is {\em rapidly and comprehensively momentum-coupled} to the strong WR wind.
The implied distance should therefore be 2.6$\pm$0.7\,kpc at which the displacement 
between successive spiral coils is 170\,AU, and the excellent match
to an Archimedian spiral confirmed.
It is very interesting to speculate that the spiral geometry, which allows a
far larger interaction between the winds and generates active shocks at much greater
radii than a non-rotating system, may also help play a role in the physics of dust
nucleation in these systems.
Indeed, due to the differing velocities of gasses in the region immediately behind
the wake, material will be piled up both at the inner and outer walls 
(see Figure~\ref{cartoon}).
Numerical simulation in three dimensions of this fascinating `twist' to a conventional 
bowshock geometry, extending the coriolis force hydrodynamic treatment of \citet{LSG07} 
to large radii, may prove a valuable guide for further progress in understanding 
these structures.


\section{The Potential Role of Radiative Braking in WR~104}
\label{radbrak}
%
%
%

For WR+OB binary systems in which the momentum of the Wolf-Rayet wind
substantially exceeds that of the OB star, the WR wind can penetrate very
deeply into the acceleration region of the OB-star wind, and in such
cases it becomes relevant to consider the role of the OB-star
{\em light} in providing the momentum balance against the WR wind.
The analysis by Gayley, Owocki, \& Cranmer (1997, hereafter GOC)
suggests that this can lead to a ``sudden radiative braking'' of
the WR wind that 
can hold the stagnation point along the line of centers farther
off from the OB star.
This could also modify the angle
for the global wind interaction
cone, with ramifications for the observations of dust formation.
GOC further provide scaling relations to identify whether such
radiative braking is likely to be of importance for any given system,
and even provide a diagram (their figure 5) to classify several of the
known close WR+O binary systems.
Let us now apply these criterion to ascertain the potential
role of radiative braking for WR~104.

Note first that the systems identified by GOC as likely to have
radiative braking are mostly relatively close binaries with periods on the
order of days or weeks, and separations of only a few OB-star radii.
For wider systems, spherical expansion dilutes the WR wind more strongly
in the region around the OB-star, making it easier for the OB-star wind
to maintain a standard wind-wind ram pressure balance.
But if the WR/OB wind momentum ratio,
\beq
P_{WR/OB} \ \equiv \ \frac{{\dot M}_{WR} V_{WR}}{{\dot M}_{OB} V_{OB}}
\, ,
\label{pwrodef}
\eeq
is sufficiently high, then even in a relatively wide system, such
as WR~104,
the standard ram pressure balance could potentially be supplanted by radiative
braking.
The model wind parameters for WR~104, from Section~\ref{basicprop}, do
indeed imply an
extreme momentum ratio, $ P_{WR/OB} \cong 305$, and so even the
relatively wide separation of $D \cong 2.4$~AU,
or $d \equiv D/R_{OB} \cong 50$~ OB-star radii,
might still induce radiative braking prior to achieving the hydrodynamical
stagnation of the winds.

\subsection{Estimating the braking radius in WR~104}

The following discussion makes use of a number of parameters that
are described in greater detail in GOC; for convenience here we 
provide a brief glossary introducing significant quantities.
Radiative acceleration is assumed proportional to $\eta$ which is
found from the effective opacity due to the integral of the forest
of spectral lines in the wind. 
The parameter $\alpha$ was introduced by \citet{CAK} to describe how
sensitively the line force increases with wind acceleration.
GOC found it convenient to define normalized quantities for
radiative braking separation $d_{rb}$ and momentum ratio $P_{rb}$ 
which could be used to scale the actual separation and momentum
ratio to gauge the significance of braking in individual cases
(see GOC for detailed derivations).

To locate WR~104 on the GOC figure 5 classification diagram,
this wind momentum ratio $P_{WR/OB}$ and separation $d$ must be
further scaled by associated
radiative braking values, $P_{rb}$ and $d_{rb}$.
To determine those, we first need to evaluate the parameter $\eta$,
given by GOC eqn. (29), which unfortunately includes an undefined
parameter $v_{\ast}$ (for that parameter, see Gayley, Owocki, \& Cranmer
1996).
Reconstruction of the GOC analysis leads to 
\beq
\eta \ = \ \frac{1}{\alpha^{\alpha/(1-\alpha)} \, (1+\alpha)}
\left ( \frac{L_{OB}}{L_{WR}} \right )^{1/(1-\alpha)} \,
\frac {2 GM_{WR}}{V_{WR}^{2} R_{OB}}
\ = \ \frac{4}{3} \, \left ( \frac{L_{OB}}{L_{WR}} \right )^{2} \,
\left ( \frac{V_{esc,WR}}{V_{WR}} \right )^{2} \, \frac{R_{WR}}{R_{OB}}
\, ,
\label{etadef}
\eeq
where the latter equality applies to the simple case that the
power index $\alpha = 1/2$, and
also recasts the scaling in terms of the ratio of WR wind speed to
surface escape speed.
Note that, in addition to replacing the undefined $v_{\ast}$, this
formulation has the advantage of being manifestly independent of the
separation $D$, as required by GOC eqns. (11), (12), and (17).

Applying the wind parameters from Section~\ref{basicprop}, we find 
(assuming $\alpha=1/2$) that for WR~104,
$\eta \cong 1.65$.
Plugging this into GOC eqn. (30) then leads to $d_{rb} \cong 2.1$.
For a standard velocity index $\beta =1$, we then further find
$P_{rb} \cong 0.642$.
Using these with the above values for $P_{WR/OB}$ and $d$,
we then find a scaled momentum ratio
\beq
{\hat P} \ \equiv \ \frac{P_{WR/OB}}{P_{rb}} \ = \ \frac{305}{0.642} \ = \ 475
\, ,
\eeq
and scaled separation
\beq
{\hat d} \ \equiv \ \frac{d}{d_{rb}} \ = \ \frac{50}{2.1} = 24
\, .
\eeq
In the $\log {\hat P}$ vs. $\log {\hat d} $ plane, systems with
radiative braking are defined by the triangular region with
both $\log {\hat d} >  0$ and $\log {\hat P} > 2 \log {\hat d}$.
The former ensures that the momentum of the OB-star light is sufficient
to keep the WR wind from impacting the OB-star surface,
and is easily satisfied in this case.
The latter condition implies that there can be no normal ram pressure
balance between the winds, and thus that radiative braking must
stop the WR wind;
since for the above parameters, we have ${\hat P} = 475 > 407 = {\hat
d}^{2}$, implying that this is condition is also satisfied, although
more marginally.
If, for example, the estimated wind momentum ratio were reduced by
more than about 15\%, or the separation were increased by more than
about 7\%, then the system would no longer satisfy this latter
condition.
In those cases, a normal ram-pressure balance is still allowed, but
even then radiative braking can play a role if the braking radius is
outside the wind-wind balance radius.

From GOC eqn. (18), we find that the braking radius for
the above standard parameter set is
\beq
r_{b} \ = \  x_{b} D \ = \  \frac{D}{1+ (d/\eta)^{1/3}}
\ = \ 0.254  \, D
\ = \ 12.1 \, R_{OB}
\, .
\label{rbeq}
\eeq
For comparison, for the simple case of constant flow speed
(i.e. velocity power index $\beta  = 0$), ram pressure
balance in the absence of direct radiative forces occurs
at a radius (cf. GOC eqn. (2))
\beq
r_{s} \ = \  x_{s} D \  =  \
\frac{D}{1 + \sqrt{P_{WR/OB}}}
\ = \ 0.054 ~ D
\ = \  2.6 \, R_{OB}
\, .
\label{rseq}
\eeq

For such a purely hydrodynamical balance, a global momentum analysis
(Canto, Raga, and Wilkin 1996;
Gayley \& Owocki in preparation)
yields a transcendental relation between the cone-opening half-angle
$\theta_{s}$ and the momentum ratio $P_{WR/OB}$,
\beq
\tan \theta_{s} - \theta_{s} \  =  \ \frac{\pi}{P_{WR/OB} - 1}
\, .
\eeq
For large momentum ratio $P_{WR/OB} \gg 1$,
use of a small-angle approximation for 
$\tan  \theta_{s} \cong \theta_{s} + \theta_{s}^{3}/3 $
gives the asymptotic form,
\beq
\theta_{s} \ \cong \ \left ( \frac{3 \pi}{P_{WR/OB}} \right )^{1/3}
~~ ; ~~ P_{WR/OB} \gg 1 
\, .
\eeq
To recover 
$\theta_{s} = 90^{o}$ for the symmetric 
momentum case $P_{WR/OB}=1$, we can write a more generally applicable
approximate solution, accurate to within about 6\% for all $P_{WR/O}$,
\beq
\theta_{s} \ \cong \ \frac{121}{31/90 + P_{WR/OB}^{1/3}}
\, 
\eeq
where using $121 = (180/\pi ) \left ( 3 \pi \right )^{1/3}$
gives this angle in degrees.

For our estimate of $P_{WR/OB} = 305 $ for WR~104, we then obtain
$
\theta_{s} \cong  18^{o}
\,,
$
in good agreement with the value from Section~\ref{basicprop} derived 
from the heuristic formula of \citet{EU93}.

\subsection{Radiative force influence on the shock-cone angle}

An important question is then, by how much can the effects of radiative
forces from the OB star potentially widen this shock cone?
Only a complete simulation can answer this definitively, but we can 
make conceptual progress using heuristic approximations to attempt to
better constrain the circumstances under which we should expect an
important radiative modification of the bow shock geometry.
Two approaches that borrow from the existing sudden radiative braking
analysis can be termed the ``Global Momentum Augmentation'' (GMA)
and the ``Radiative Wall'' (RW) approach.
Each gives similar results for the opening angle, and so here we
briefly outline just the former, deferring to Gayley \& Owocki 
(in preparation) for a detailed discussion of both approaches.


In the GMA approach, the radiation interaction
with the incident WR wind is assumed to augment the OB-star
wind momentum flux by a globally constant factor, 
set by the requiring that the stagnation point of the combined effective 
momentum fluxes equal the braking radius in eqn. (\ref{rbeq}).
We write this factor as $\gamma \tau$,
where $\gamma$ is the ratio of the OB-star radiative momentum flux to
its wind momentum flux, $L_{OB}/{\dot M}_{OB} v_{OB} c$,
and $\tau$ is an optical-depth-like parameter that
simulates the coupling of the OB radiative momentum flux to the WR wind.
In analogy to the global analysis cited above (Gayley \& Owocki, in 
preparation), we find for the large momentum ratio limit, 
\beq
\theta_s \ \cong \ 121 \, 
\left ( \frac{1 \ + \ \gamma \tau}{P_{WR/OB}} \right )^{1/3} \ .
\eeq

In this scenario radiative braking would alter
the shock geometry whenever $\gamma \tau$ is appreciable, and since
$\gamma$ in WR~104 is expected to be large, this is not a terribly
restrictive requirement for the coupling parameter $\tau$.
Since the  braking radius $r_{rb}$ scales roughly with $\eta^{-1/3}$
when braking is important, the condition that the effective
momentum ratio used in $r_s$ 
be reduced enough to achieve 
$r_{rb} = r_s$ (see eqs. [\ref{rbeq}] and [\ref{rseq}])
is tantamount to requiring
\beq
\gamma \tau \ \cong \ \left ( \frac{ \eta  R_{OB}}{D} \right )^{2/3}
P_{WR/OB} \ .
\eeq
This implies a shock opening angle 
\beq
\theta_{s} \ \cong \ 121 \, 
\left ( \frac{ \eta  R_OB}{D} \right )^{2/9}
\, ,
\eeq
which for $\eta = 1.65$ and $D/R_{OB} = 46.5$ evaluates to $\theta_{s}
\cong 58^{o}$, representing a factor $\sim 3.2$ increase over the
purely hydrodynamic value $\theta_{s} \cong 18^{o}$.
This may be a significant overestimate, because of the optimistic
assumptions made about the radiative momentum coupling, but
it demonstrates that radiative braking could indeed increase the shock cone
opening if the WR wind opacity continues to be high throughout the flow.

A similar result can be reached through the radiative wall approach.
Although both approximations are too heuristic to 
be considered quantitatively reliable,
they do suggest that strong radiative braking for the assumed parameters
in WR~104 should be expected to widen the bow shock angle considerably,
possibly by a factor of 2 or more when compared to
the purely hydrodynamic result
$\theta_{s} \cong 18^{o}$, as long as the incident WR wind maintains
the same high levels of opacity for the OB starlight that it must have
had to have been radiatively driven from the WR star.
This assumption is currently of unknown validity, and indeed
if we take the opposite limit, and say that the line-opacity is just
what is needed to drive the relatively weak OB-star wind, then
radiative braking will be {\it much} weaker and would not likely alter
the shock-cone geometry in any measurable way.

To address intermediate values of the effective WR wind opacity,
recalling that $\eta$, for fixed $\alpha$, is simply proportional to
the integrated spectrum-weighted line opacity
distribution with no regard
to optical depth (i.e., self-shadowing) corrections.
Thus one may either make assumptions about this line opacity
to predict $\eta$, or one may use observations of the bow shock cone
to infer $\eta$ and reason to conclusions about the line distribution.
Further analysis will be required to constrain these expectations
more quantitatively, but it seems at least plausible that
radiative braking may be important in the wind-wind 
interaction and shock-cone geometry of WR~104, {\em
if} the line opacity of the WR wind interacting with the OB starlight 
is substantially enhanced above what is needed to drive the OB-star wind itself.
Reversing the logic, it may be concluded that evidence for a significant
widening of the bow-shock cone relative to what may be explained 
hydrodynamically may be interpreted as evidence that the WR wind 
does indeed carry with it an intrinsically high opacity when it
impinges into the region of strong OB-star radiative flux.

Our images of the WR~104 plume are able to rule out some of the more 
extreme enlargements to the cone opening angle in the above scenarios.
Fits to the data by HMSK found that model cone half-angles of
$40^\circ$ or greater (twice the hydrostatic value) were rejected.
With still higher angular resolutions than those obtained here,
it should be possible to get a direct measurement of the cone
opening angle from the images, revealing the role played by 
radiative braking on the shock geometry.

\section{The Profile of the Spiral Plume}

\begin{figure}
\centerline{\includegraphics[angle=180,scale=.8]{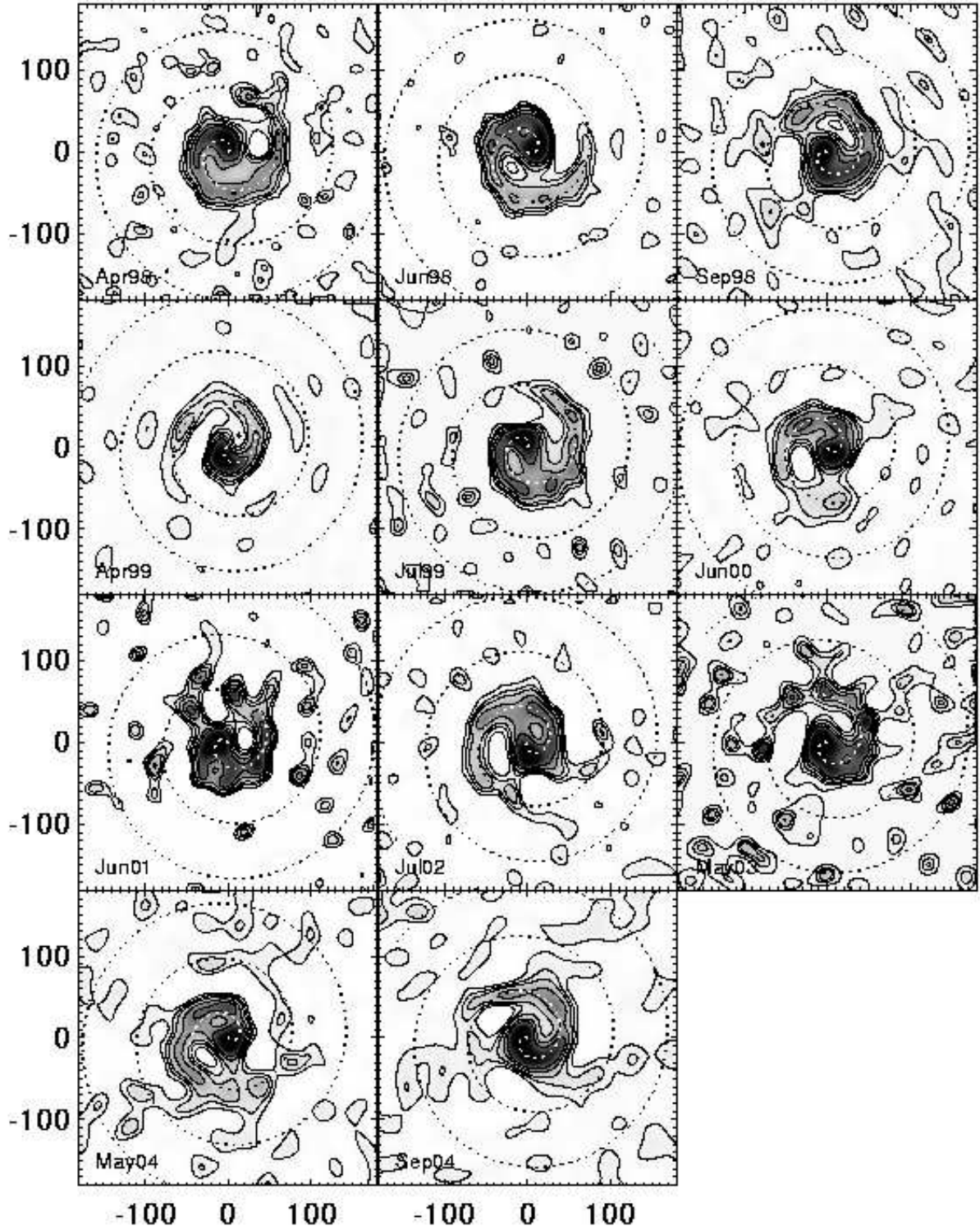}}
\caption{\label{hmaps}
Images recovered from data in the H filter over 11 observing epochs.
See caption to Figure~\ref{kmaps} for details.
}
\end{figure}

\begin{figure}
\centerline{\includegraphics[angle=180,scale=.8]{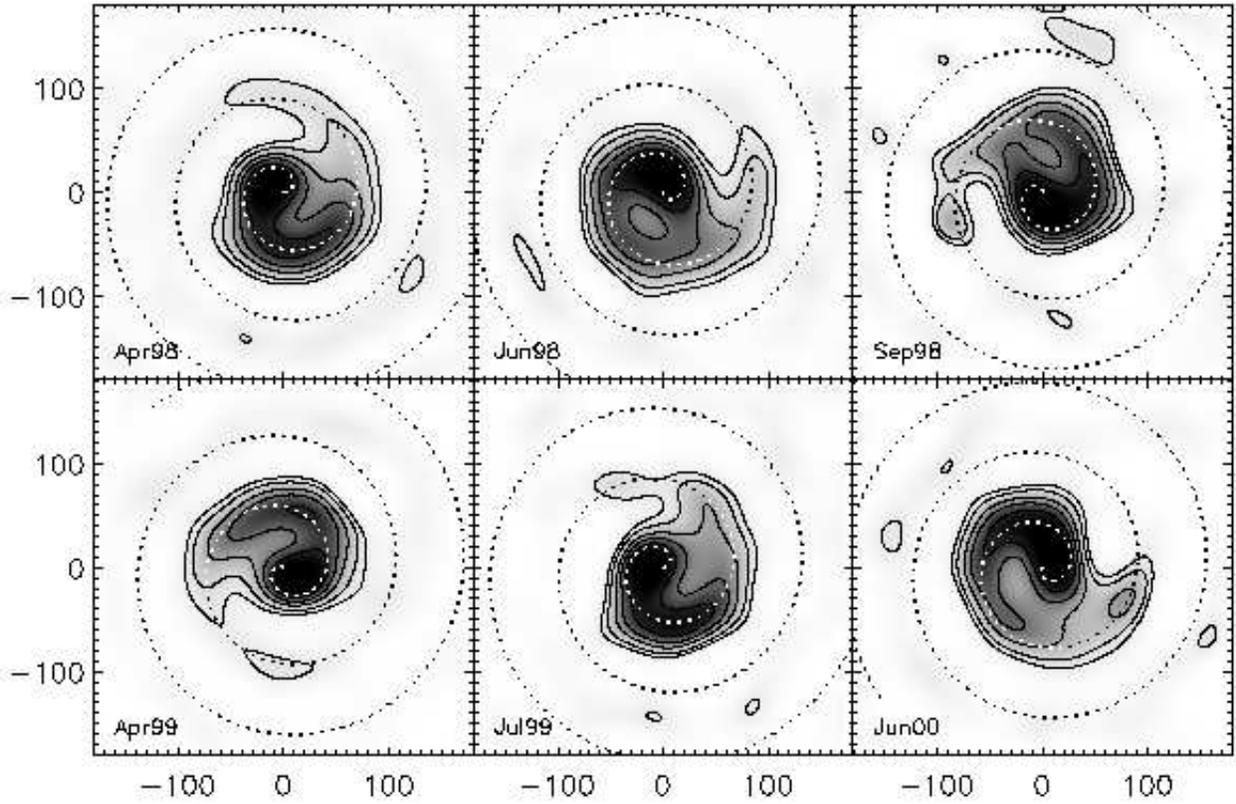}}
\caption{\label{lmaps}
Images recovered from data in the PAHCS filter over 6 observing epochs.
Contour levels are 1, 2, 5, 10, 20\,\% of the peak. Otherwise
similar to Figure~\ref{kmaps}. 
}
\end{figure}

\begin{figure}
\centerline{\includegraphics[angle=0,scale=1.0]{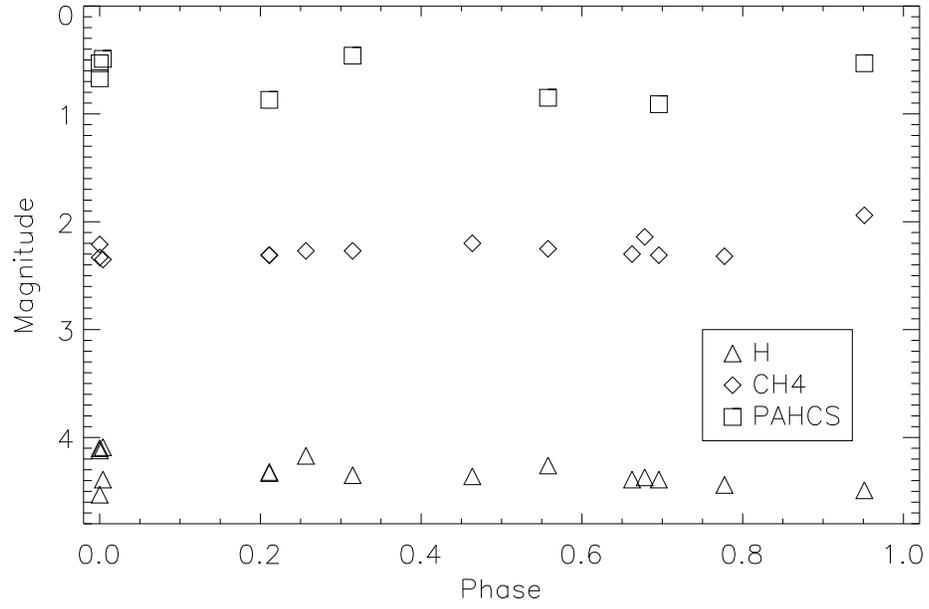}}
\caption{\label{lightcurves}
Photometry data for WR~104 obtained in our three filter bandpasses
(key inset), folded with the 241.5\,d period.
}
\end{figure}

\begin{figure}
\centerline{\includegraphics[angle=0,scale=0.5]{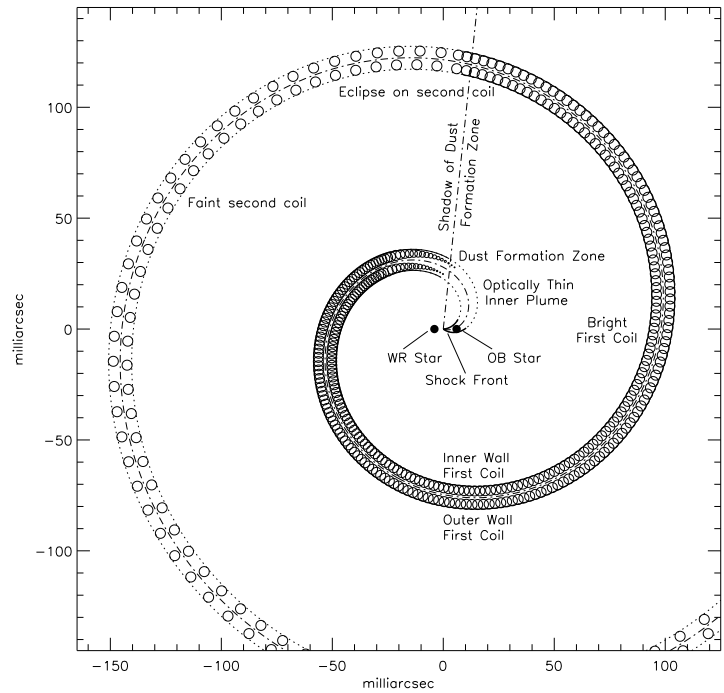}}
\caption{\label{cartoon}
Schematic diagram of the key elements in the WR~104 system.
The geometry depicted is that of a face-on view onto a 
$20^\circ$ cone half-angle pinwheel. 
The axis of the system, and the line of shadow cast at $84^\circ$ 
to the spiral origin ($\theta=0$), are 
indicated as dot-dashed lines. Key elements are labeled, and 
some, such as the binary separation, are not to scale 
(depicted here about a factor of ten too large).
}
\end{figure}

\begin{figure}
\centerline{\includegraphics[angle=0,scale=1.0]{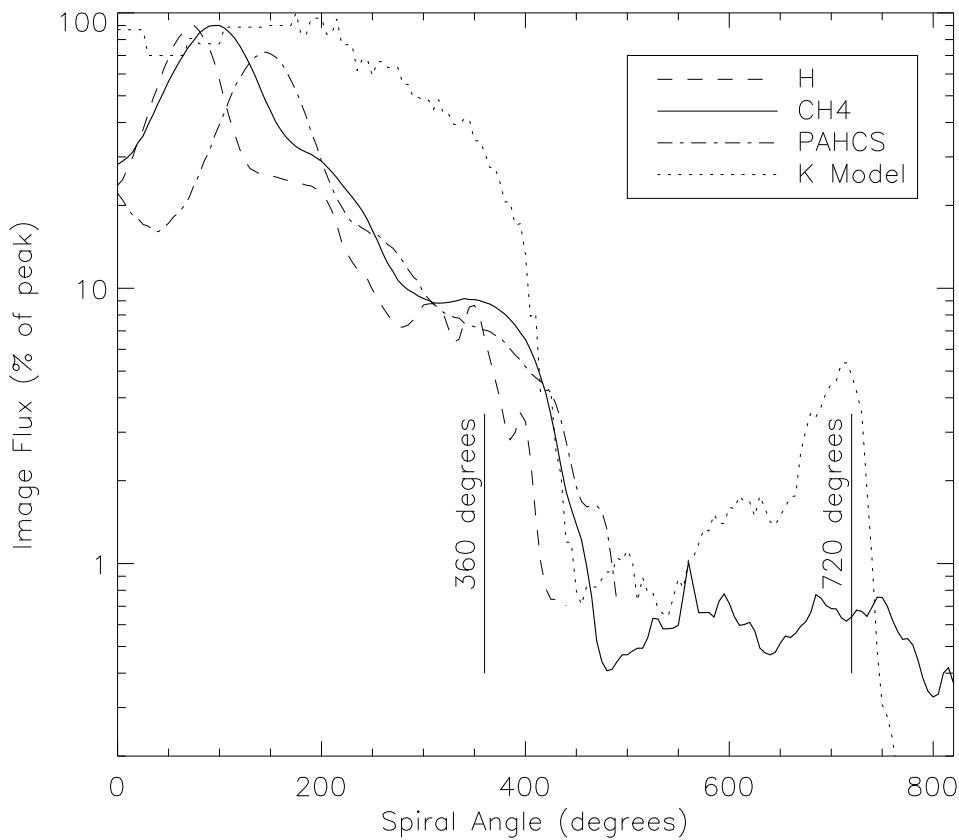}}
\caption{\label{slice}
Average brightness profile of the plume as a function of angular displacement 
along the spiral model locus.
Data for our three filter bandpasses are given (linetypes indicated in the key)
together with a profile extracted from the numerical radiative transfer 
simulations of \citet{Harries04} specific to our CH4 filter in the K band.
Data for H and PAHCS are terminated shortly after the first full
turn, at which point the signal-to-noise in the images is insufficient to 
discern real structure from noise.
For convenience, intervals of 1 and 2 complete revolutions are marked.
}
\end{figure}

Using the excellent fits provided by the theoretical spiral locus to the observed
plume of emission from Figures~\ref{kmaps},~\ref{hmaps}~\&~\ref{lmaps}, an 
investigation of the brightness profile along the length of the dust plume was
performed.
The flux, summed over a band following the model spiral, was accumulated over all 
images recovered. 
Average profiles, plotted as a function of angular displacement along the 
spiral, are given in Figure~\ref{slice} for all three filter bandpasses.
Also overplotted is the same profile along the spiral calculated from the 
synthetic radiative-transfer images of HMSK.
All experimental curves follow a similar form (with some departures for PAHCS 
which we attribute to the systematically lower resolution) exhibiting a steep 
rise in flux from the spiral center to a point about $90^\circ$ downstream 
followed by a rapid drop as the bright core of the system is traversed. 
There now follows a region of roughly monotonic decline between about 
$150-400^\circ$, at which point there is a very sharp drop in brightness 
flattening to level of less than 1\% of the peak for the second coil 
(CH4 filter only).

The initial sharp rise in flux seen in Figure~\ref{slice} gives another way to
visualize the ``standoff distance'' between the spiral center and brightest pixel,
measured earlier to be $13.3\pm1.4$\,mas or $85^\circ\pm14^\circ$ downstream along 
the spiral.
Given the 241.5\,d period, this corresponds to a kinematic age for the
material of $57\pm10$\,d.
This value is in remarkable agreement with the value of $\sim$\,58--87\,d for 
the delay between periastron passage and the peak in the dust production in
the well-studied episodic colliding-wind binary dust producer WR~140 
\citep{Williams90,Marchenko03}.
The finding that the peak in dust production occurs at some distance
downstream from the colliding-wind stagnation point ($\sim$30\,AU for WR~104)
points to a more complicated picture for dust formation physics in 
colliding-wind binaries than the simple shock compression/cooling models
\citep{Usov91}.
A number of avenues for additional physics have been presented 
\citep[see also][]{Crowther03}, with one likely element being escape from the 
dust-hostile immediate circumstellar radiation environment.
Chemical seeding of the WR wind (predominantly helium) with hydrogen from 
the OB~star at the contact surface may help to enhance dust formation 
pathways \citep{LT02}, although such mixing may not be very rapid. 
Shielding from the intense radiation field does appear to be a key ingredient,
and sets the playing field for sharply nonlinear grain growth in which small
clumps may self-shadow larger areas downstream in a runaway process. 
Recent numerical simulations of colliding flows have shown that matter is 
squeezed into thin high-density shells threaded by still higher density 
filaments as a result of supersonic turbulence \citep{WF00}.
It was suggested these conditions can persist for a considerable time in the
flow before dissipation by internal shocks and vortex cascades.
Further to these earlier ideas, we now also add the idea of wind entanglement
which can generate significant shock activity well downstream of the nose,
as discussed in earlier sections.
Whatever mix of these or other elements may go into the next generation of
colliding-wind dust formation models, there certainly seems scope to address
the outstanding issue of whether the dusty-WC phenomena has a necessary link
to binarity \citep{Monnier99,WV00,Monnier07}, or whether novel dust formation
pathways in an isolated wind may also be important in some circumstances
\citep{Zubko98,Cherchneff00}.

After the peak at $85^\circ$, the profiles of Figure~\ref{slice} exhibit a
steady decline (but note the logarithmic y-axis).
This is in accord with the relatively flat temperature profile found in
models of the centrally-illuminated dust spiral models of HMSK,
although the bright peak near the heart of the system is not reproduced
in the simulations.
A fairly detailed comparison between the radiative transfer results and 
published masking images of WR~104 can be found in HMSK, 
although our new results here highlight some discrepancies which we hope 
will motivate future refinements.
The flux level of the second winding of the spiral is well reproduced
at the beginning ($480^\circ$--$600^\circ$) but the secondary bright knot 
in the model tail seen at 2 complete turns was not observed in the data.
Although a detailed comparison is beyond the scope of the present paper,
one possible implication is that the optical depth of the dust in the
model is not high enough in the core of the system. 
Increasing it may help to generate the bright peak at $85^\circ$, and
at the same time decrease the illumination on the second coil therefore
suppressing the spurious feature at $720^\circ$.

The very sharp decline in flux between 400--$480^\circ$ is a result of the
shadow cast by the first winding of the spiral onto the second, causing 
a rapid drop in temperature for the eclipsed material.
This feature has been commented upon since the discovery of the pinwheels
\citep{nature99}, and was also reproduced in radiative transfer models
(HMSK).
What is particularly interesting is that the terminator is found at an
angular ordinate of $\sim 440^\circ$ along the spiral, or about $80^\circ$
beyond one complete revolution.
This extra angular displacement (easily visualized in the figure as the
full revolution $360^\circ$ location has been marked) gives further 
independent confirmation for the existence and size of the optically thin 
inner cavity surrounding the central stars.
We refer the reader back to Figure~\ref{cartoon} for an easily understood
and intuitive depiction of the angular locations of the inner optically
thin plume, and the geometry of the shadow cast upon the second coil by
the first.

Also of interest in helping to constrain future generations of dust plume
models are the multi-wavelength observations of the spiral given in
Figures~\ref{kmaps},~\ref{hmaps}~\&~\ref{lmaps}. 
In particular, there was a systematic rotational displacement found with 
observing wavelength with a (lower limit) position angle of $6^\circ$ from 
H--CH4 and $4^\circ$ from CH4--PAHCS (true values may be a factor of $\sim$2
larger -- see discussion in Section~\ref{HLimages}).
This implies that the transverse temperature profile of the plume 
can be explored by our high resolution techniques.
Short wavelength observations will preferentially weight the hottest material 
on the inner edge of the plume facing the central stars, while longer wavelengths
will give a better indicator of the bulk concentration of warm dust. 
Again, we must defer detailed modelling of these effects to a future paper,
but note that preferential radiative heating of the inner wall was predicted 
in models of HMSK.
An alternate scenario is that of asymmetric heating by shocks of differing 
strength at the inner and outer walls, as recently demonstrated in the rotating 
hydrodynamic simulations of \citet{LSG07}.

\section{Geometrical Implications of the Plume Structure}

A key finding from earlier studies of the WR~104 pinwheel has been the
low eccentricity of the orbit of the central stars \citep{nature99,wrconf02}.
As noted earlier \citep{Monnier99}, this finding may be a ``smoking gun''
pointing to previous episodes of tidal circularization and likely
Roche-lobe interactions with possible mass transfer and/or envelope 
stripping. 
Such an event provides one of the two possible pathways to the creation of
a Wolf-Rayet \citep{Paczynski67}; the other being strong wind-driven mass 
loss in a single star \citep{Chiosi78}.
With our long time-sequence of observations sampling all phases of the orbit, 
we are able to put tight constraints on the allowed levels of orbital 
eccentricity in the case of WR~104.
Although it would be possible to fit eccentric spiral models directly to the
data, we favored a more indirect approach as for small to moderate levels of 
eccentricity, the departure in shape from an Archimedian spiral was subtle.
The most easily detectable effect that an eccentric orbit has in perturbing 
the rotating spiral model away from its ideal form will be in 
departures from a constant rate of rotation.

\begin{figure}
\centerline{\includegraphics[angle=0,scale=1.0]{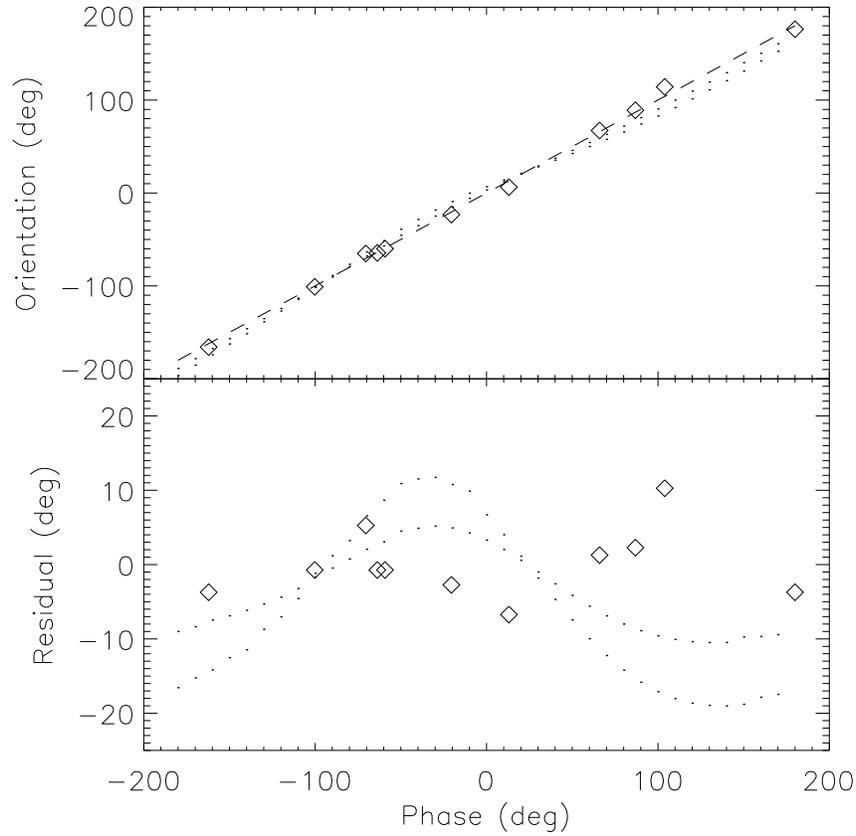}}
\caption{\label{rotfig}
(Upper panel) The rotation of the pinwheel at each observing epoch
from fits to data in Figure~\ref{kmaps}, plotted as a function of the 
orbital phase. 
The dashed line gives a simple model of uniform angular velocity.
(Lower panel) the residual between the data and the uniform model.
Both panels have overplotted dotted lines giving examples of expected
curves for elliptical orbits with eccentricity $e = 0.1$ and 0.2.
}
\end{figure}

We have therefore performed a careful fit of the best-fit orientation of
the spiral structure at each epoch, and compared the values obtained with
the model of a simple, constant angular velocity. 
This was done in two different ways which gave similar results: firstly by 
fitting to points extracted from the crest of the bright spiral ridge, and 
secondly by fitting each full image to a mean template.
The uniformity of the angular rotation rate extracted from the real data, 
which is given in in Figure~\ref{rotfig}, was then compared with results from 
model simulations.
A family of synthetically-generated spiral images of varying eccentricity 
were fit in an identical fashion to the real data, allowing us to derive
quantitative limits on the levels of eccentricity present in the WR~104 orbit.
Figure~\ref{rotfig} shows the rotational uniformity from the data and examples
from two of the synthetic eccentric model images. 
The rotation is seen to be highly uniform, ruling out all but very modest 
eccentricities.
A 2$\sigma$ upper limit of $e$~\simle~0.06 was obtained in this fashion.

\begin{figure}
\centerline{\includegraphics[angle=90,scale=0.5]{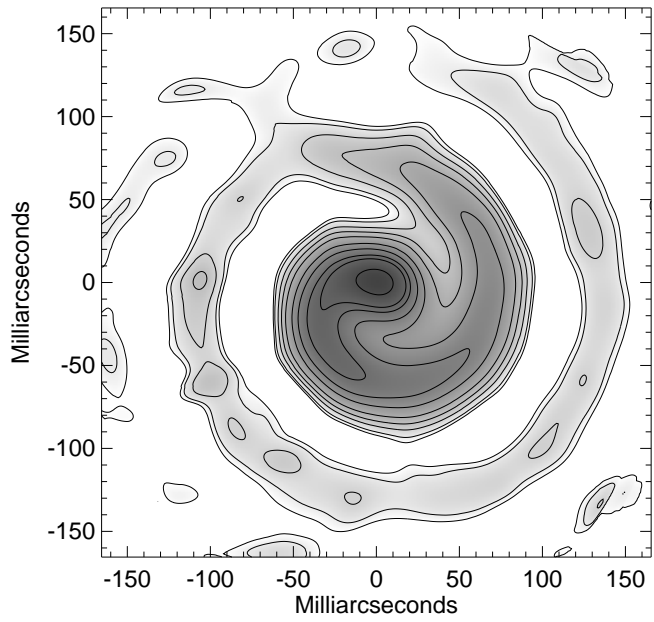}}
\caption{\label{kstack}
Stacked composite image of all epochs of data in the CH4 filter.
Contour levels are .1, .2, .5, 1, 2, 5, 10, 20, 30, 70\,\% of the peak. 
Individual images have been projected to a face-on viewing angle
and derotated to the first Apr98 epoch according to the best-fit
Archimedian spiral model prior to coadding. 
}
\end{figure}

Having thus established the properties of the best-fit uniform Archimedian 
spiral model, it was then possible to generate a stacked composite image
averaging over all epochs of data taken in the CH4 filter.
Individual data frames were corrected for inclination on the sky (simulating
a perfectly face-on view) and de-rotated to coincide with the first (Apr98)
epoch before being co-added. 
A composite image, displaying significantly greater dynamic range than any
of the single-epoch images from Figure~\ref{kmaps}, is given in Figure~\ref{kstack}.
Variations in morphology between different epochs beyond those expected from 
the rotating spiral model are minor, and all appear to be within the noise level
of the image recovery process. 
We therefore hope that this image, as a mean over all epochs, may provide
a resource to modellers and others interested in the best possible snapshot
of the geometry of these pinwheel systems.

The flat infrared lightcurves of Figure~\ref{lightcurves} discussed above 
affirm the status of WR~104 as a ``constant'' dust-producer \citep{Williams97}, 
yet stand in contrast to the large 2.7\,mag ``quasi-periodic'' variability 
(periodic component \simge~1\,mag) in the visible lightcurves reported by 
\citet{Kato02}.
With a narrow cone half-angle ($20^\circ$; HMSK), face-on
orientation to the line of sight and very low levels of orbital eccentricity, 
the spiral dust plume gives no obvious mechanism for modulating the visible
flux from this system.
The lack of detectable infrared variation argues for steady and constant dust
production at all orbital phases, with minimal effects from any inclination
to the the line of sight.
Visible light scattered from the dust plume may be more strongly modulated
by rotation and inclination effects, however it is difficult to see how
variations in the V lightcurve as large as \simge~1\,mag could arise from this.
Possible scenarios include opacity effects from gas (or possibly small quantities
of dust) associated with the innermost regions of the shock cone which modulate
the light of the OB~star, line-of-sight (polar-axis) dust creation independent of 
the colliding wind mechanism, or some geometrical effect from a sharp opacity 
gradient possibly created at an earlier phase 
\citep[further discussion can be found in][]{Kato02}.

\section{WR Systems as Supernova Progenitors}

As a final point of discussion, we also explore the implications of future 
evolution of this system with particular regard to the face-on viewing angle 
to Earth.
The formal Archmiedian spiral fits yielded an inclination $i = 12^\circ$, but
for such a small angle to the line-of-sight the uncertainty was large and fully
consistent with the range from $0^\circ$~--~$16^\circ$.
A further source of uncertainty arises from the limitations of the simple 
two-dimensional spiral model used.
In reality, the pinwheel is of course a three-dimensional structure which may
present us with more complicated opacity effects such as limb-brightening.
If we consider a single circular cross-section through the conical shock, as 
this is carried downstream by the WR wind, it will become inflated and distorted 
by the spherical expansion into an elongated ``D''.
Eventually at large radii, the plume shape will resemble an equatorial band
wrapped around a sphere, subtending an angle equal to that of the shock cone. 
This underlying spherical geometry gives rise to a fundamental insensitivity 
to the line-of-sight inclination: modest changes of tilt for such a structure 
result in only a very small or zero change in observed aspect ratio.
For these reasons, our upper limit of $16^\circ$ may be an underestimate.
Firmer constraint of the inclination from the data presented here would require
detailed radiative transfer modelling with varying lines of sight onto a
three-dimensional structure.

Despite these quantitative caveats, it remains apparent that in WR~104, Earth
has an approximately pole-on view onto a relatively nearby, massive WR binary 
system.
Given the prehistory of mass transfer and envelope-stripping suggested above, 
it therefore seems likely that Earth also lies in the polar direction with
respect to the spin axis of both WR and OB stars due to the prior angular
momentum evolution of such binary systems \citep[e.g.][]{Zahn77,Tassoul87}.
As both objects in the system should eventually explode as core-collapse
supernovae, it is interesting to speculate on how our ``privileged'' polar
vantage point may affect our experience of this event, which should happen
within a timescale of several hundred thousand years for the WR component
\citep{Maeder81,CM86}.
It is becoming more accepted that all supernova explosions probably exhibit
some preferred axis aligned with the progenitor stellar spin.
As one of the closest imminent WC-class Wolf-Rayets, an isotropic supernova 
in WR~104 at kiloparsec scale distances would no doubt put on an impressive 
show, but any impact on Earth's biosphere is likely to be negligible \citep{ES95}.
However for a highly {\em anisotropic} explosion, the most extreme example of 
which would be a Gamma Ray Burst, effects could be significant if Earth
lies within the \simle~12$^\circ$ opening angle \citep{Frail01} of the
burst, even at 2\,kpc distances \citep{Melott04,Thomas05a,Thomas05b}.
 

Could WR~104 produce a GRB?
Certainly a case could be made that the OB component of the binary 
system, which was the recipient of mass and angular-momentum transfer in the 
prior Roche-lobe overflow event, may meet the requirements for envelope
mass and angular momentum for many current models of GRB formation
\citep[see][for a discussion of WR binaries as GRB progenitors]{Petrovic05}.
The Wolf-Rayet component, with WC spectral type, has arguably the shortest
fuse, being in the last known stable phase before supernova.
Although prior stripping may have lowered the envelope mass into a regime
less favorable for conventional collapsar GRB's, there are also 
more exotic magnetar (millisec pulsar driven) GRB models which may prove
viable \citep{Gaensler05,Metzger07}.
In short, this is a very active and rapidly-evolving field, and the 
uncertainties outnumber the firm conclusions in arriving at a 
mass/metallicity/evolutionary portrait of likely GRB progenitors.
Whether or not the WR~104 system will play host to a future GRB, it
does seem clear that two pole-oriented supernovae will occur, and that
energy and matter will be preferentially ejected along this axis
\citep[see][for a discussion of the effects of stellar rotation on
SN explosions]{FW04}.
Further observations to constrain the inclination of the orbital plane, 
which should be possible with spectroscopic monitoring, are therefore
encouraged.


\acknowledgments

The data presented herein were obtained at the W.M. Keck Observatory, 
which is operated as a scientific partnership among the California Institute 
of Technology, the University of California and the National Aeronautics and Space
Administration.  The Observatory was made possible by the generous
financial support of the W.M. Keck Foundation.
The Authors would like to particularly thank the instrument technicians and
engineering crew at the observatory for their help and support of our unorthodox 
observing techniques.
Devinder Sivia kindly provided the maximum entropy mapping program ``VLBMEM'',
which we have used to reconstruct our diffraction limited images.
This work has been supported by grants from the National Science Foundation
and the Australian Research Council. 
We thank Charles Townes for his long-standing support of this program, 
and Bryan Gaensler, Shami Chatterjee and Grant Hill for helpful discussions.

\clearpage

\clearpage

\clearpage

\clearpage

\clearpage

\clearpage

\clearpage

\clearpage

\clearpage

\clearpage

\begin{deluxetable}{lcl}
\tablewidth{0pt}
\tablecaption{Journal of Interferometric Observations.}
\tablehead{
\colhead{Date} & JD-2450000 &
\colhead{Filter} } 
\startdata
1998 Apr 14 & ~918 & H,CH4,PAHCS \\
1998 Apr 15 & ~919 & H,CH4,PAHCS \\
1998 Jun 04 & ~969 & H,CH4,PAHCS \\
1998 Sep 29 & 1086 & H,CH4,PAHCS \\
1999 Apr 25 & 1294 & H,CH4,PAHCS \\
1999 Jul 29 & 1389 & H,CH4,PAHCS \\
2000 Jun 23 & 1719 & H,CH4,PAHCS \\
2001 Jun 11 & 2072 & H,CH4       \\
2002 Jul 23 & 2479 & H,CH4       \\
2003 May 12 & 2772 & H,CH4       \\
2004 May 28 & 3154 & H,CH4       \\
2004 Sep 03 & 3154 & H,CH4       \\
\enddata
\label{obstable}
\end{deluxetable}

\begin{deluxetable}{ccc}
\tablewidth{0pt}
\tablecaption{Properties of Filters}
\tablehead{
\colhead{Name} & \colhead{Center} & \colhead{Bandwidth} \\
 &  \colhead{($\mu$m)} & \colhead{($\mu$m)}  \\
}
\startdata
H      & 1.656 & 0.333 \\
CH4    & 2.269 & 0.155 \\
PAHCS  & 3.083 & 0.101 \\
\enddata
\label{filtable}
\end{deluxetable}


\begin{thebibliography}{}

\bibitem[Bethe(1990)]{Bethe90} 
Bethe, H.~A.\ 1990, Reviews of Modern Physics, 62, 801 

\bibitem[Canto et al.(1996)]{Canto96} 
Canto, J., Raga, A.~C., \& Wilkin, F.~P.\ 1996, \apj, 469, 729 

\bibitem[Castor et al.(1975)]{CAK} 
Castor, J.~I., Abbott, D.~C., \& Klein, R.~I.\ 1975, \apj, 195, 157 

\bibitem[Cherchneff et al.(2000)]{Cherchneff00} 
Cherchneff, I., Le Teuff, Y.~H., Williams, P.~M., \& Tielens, A.~G.~G.~M.\ 
2000, \aap, 357, 572 

\bibitem[Chiosi et al.(1978)]{Chiosi78} 
Chiosi, C., Nasi, E., \& Sreenivasan, S.~R.\ 1978, \aap, 63, 103 

\bibitem[Chiosi \& Maeder(1986)]{CM86} 
Chiosi, C., \& Maeder, A.\ 1986, \araa, 24, 329 

\bibitem[Crowther(1997)]{Crowther97} 
Crowther, P.~A.\ 1997, \mnras, 290, L59 

\bibitem[Crowther(2003)]{Crowther03} 
Crowther, P.~A.\ 2003, \apss, 285, 677 

\bibitem[Danchi et al.(2001)]{mwc349} 
Danchi, W.~C., Tuthill, P.~G., \& Monnier, J.~D.\ 2001, \apj, 562, 440 

\bibitem[Eichler \& Usov(1993)]{EU93} 
Eichler, D., \& Usov, V.\ 1993, \apj, 402, 271 

\bibitem[Ellis \& Schramm(1995)]{ES95} 
Ellis, J., \& Schramm, D.~N.\ 1995, 
Proceedings of the National Academy of Science, 92, 235 

\bibitem[Frail et al.(2001)]{Frail01} 
Frail, D.~A., et al.\ 2001, \apjl, 562, L55 

\bibitem[Fryer \& Warren(2004)]{FW04} 
Fryer, C.~L., \& Warren, M.~S.\ 2004, \apj, 601, 391 

\bibitem[Gaensler et al.(2005)]{Gaensler05} 
Gaensler, B.~M., McClure-Griffiths, N.~M., Oey, M.~S., 
Haverkorn, M., Dickey, J.~M., \& Green, A.~J.\ 2005, \apjl, 620, L95 

\bibitem[Gayley et al.(1997)]{GOC97} 
Gayley, K.~G., Owocki, S.~P., \& Cranmer, S.~R.\ 1997, \apj, 475, 786 

\bibitem[Gayley et al.(1996)]{GOC96} 
Gayley, K.~G., Owocki, S.~P., \& Cranmer, S.~R.\ 1996, 
Workshop on Colliding Winds in Binary Stars to Honor Jorge Sahade, 
vol.~5, p.~68, 68 

\bibitem[Gull \& Skilling(1984)]{mem} 
Gull, S.F. and Skilling, J. 1984, Proc. IEEE,  131, 6

\bibitem[Haniff \& Buscher(1992)]{HB92}
Haniff, C.~A., \& Buscher, D.~F.\
1992, J.~Opt.~Soc.~Am.~A, 9, 203

\bibitem[Harries et al.(2004)]{Harries04} 
Harries, T.~J., Monnier, J.~D., Symington, N.~H., \& Kurosawa, R.\ 
[HMSK] 2004, \mnras, 350, 565 

\bibitem[H\"{o}gbom(1974)]{clean} 
H\"{o}gbom, J. 1974, \apjs, 15, 417

\bibitem[Howarth \& Schmutz(1992)]{HS92} 
Howarth, I.~D., \& Schmutz, W.\ 1992, \aap, 261, 503 

\bibitem[van der Hucht(2001)]{7cat} 
van der Hucht, K.~A.\ 2001, New Astronomy Review, 45, 135 

\bibitem[Ireland et al.(2006)]{macim} 
Ireland, M.~J., Monnier, J.~D., \& Thureau, N.\ 2006, \procspie, 6268,  

\bibitem[Kato et al.(2002)]{Kato02} 
Kato, T., Haseda, K., Yamaoka, H., \& Takamizawa, K.\ 
2002, \pasj, 54, L51 

\bibitem[Kuiper(1941)]{Kuiper41} 
Kuiper, G.~P.\ 1941, \apj, 93, 133 

\bibitem[Lemaster et al.(2007)]{LSG07} 
Lemaster, M.~N., Stone, J.~M., \& Gardiner, T.~A.\ 2007, \apj, 662, 582 

\bibitem[Le Teuff(2002)]{LT02} 
Le Teuff, Y.~H.\ 2002, Interacting Winds from Massive Stars, 260, 223 

\bibitem[Lundstrom \& Stenholm(1984)]{LS84} 
Lundstrom, I., \& Stenholm, B.\ 1984, \aaps, 58, 163 

\bibitem[Maeder(1981)]{Maeder81} 
Maeder, A.\ 1981, \aap, 99, 97 

\bibitem[Marchenko et al.(2002)]{Marchenko02} 
Marchenko, S.~V., Moffat, A.~F.~J., Vacca, W.~D., C{\^o}t{\'e}, S., \& Doyon, R.\ 
2002, \apjl, 565, L59 

\bibitem[Marchenko \& Moffat(2006)]{MM06} 
Marchenko, S.~V., \& Moffat, A.~F.~J.\ 2006, 
ArXiv Astrophysics e-prints, arXiv:astro-ph/0610531 

\bibitem[Matthews et al.(1996)]{nirc96}
Matthews, K., Ghez, A.~M., Weinberger, A.~J., \& Neugebauer, G.\
1996, \pasp, 108, 615

\bibitem[Mauron \& Huggins(2006)]{MH06} 
Mauron, N., \& Huggins, P.~J.\ 2006, \aap, 452, 257 

\bibitem[Marchenko et al.(2003)]{Marchenko03} 
Marchenko, S.~V., et al.\ 2003, \apj, 596, 1295 

\bibitem[Melott et al.(2004)]{Melott04}
Melott, A.~L., et al.\ 2004, International Journal of Astrobiology, 3, 55

\bibitem[Metzger et al.(2007)]{Metzger07} 
Metzger, B.~D., Thompson, T.~A., \& Quataert, E.\ 2007, 
ArXiv e-prints, 704, arXiv:0704.0675 

\bibitem[Mondal \& Chandrasekhar(2002)]{MS02} 
Mondal, S., \& Chandrasekhar, T.\ 2002, \mnras, 334, 143 

\bibitem[Monnier et al.(1999)]{Monnier99} 
Monnier, J.~D., Tuthill, P.~G., \& Danchi, W.~C.\ 1999, \apjl, 525, L97 

\bibitem[Monnier et al.(2000)]{Monnier00} 
Monnier, J.~D., Tuthill, P.~G., \& Danchi, W.~C.\ 2000, \apj, 545, 957 

\bibitem[Monnier et al.(2002)]{Monnier02} 
Monnier, J.~D., Greenhill, L.~J., Tuthill, P.~G., \& Danchi, W.~C.\ 
2002, \apj, 566, 399 

\bibitem[Monnier et al.(2002)]{Monnier02b} 
Monnier, J.~D., Tuthill, P.~G., \& Danchi, W.~C.\ 
2002, \apjl, 567, L137 

\bibitem[Monnier et al.(2004)]{Monnier04} 
Monnier, J.~D., et al.\ 2004, \apj, 605, 436 

\bibitem[Monnier et al.(2007)]{Monnier07} 
Monnier, J.~D., Tuthill, P.~G., Danchi, W.~C., Murphy, N., \& Harries, T.~J.\ 
2007, \apj, 655, 1033 

\bibitem[Moran \& Reichart(2005)]{MR05} 
Moran, J.~A., \& Reichart, D.~E.\ 2005, \apj, 632, 438 

\bibitem[Paczy{\'n}ski(1967)]{Paczynski67} 
Paczy{\'n}ski, B.\ 1967, Acta Astronomica, 17, 355 

\bibitem[Pauls et al.(2005)]{oifits05} 
Pauls, T.~A., Young, J.~S., Cotton, W.~D., \& Monnier, J.~D.\ 2005, \pasp, 117, 1255 

\bibitem[Petrovic et al.(2005)]{Petrovic05} 
Petrovic, J., Langer, N., Yoon, S.-C., \& Heger, A.\ 2005, \aap, 435, 247 
 
\bibitem[Rajagopal et al.(2004)]{Jay04} 
Rajagopal, J.~K., Barry, R., Lopez, B., Danchi, W.~C., 
Monnier, J.~D., Tuthill, P.~G., \& Townes, C.~H.\ 
2004, \procspie, 5491, 1120 

\bibitem[Rochowicz \& Niedzielski(1995)]{RN95} 
Rochowicz, K., \& Niedzielski, A.\ 1995, Acta Astronomica, 45, 307 

\bibitem[Ryder et al.(2004)]{Ryder04} 
Ryder, S.~D., Sadler, E.~M., Subrahmanyan, R., Weiler, K.~W., 
Panagia, N., \& Stockdale, C.\ 
2004, \mnras, 349, 1093 

\bibitem[Sivia(1987)]{devinder} 
Sivia, D.S. 1987, Ph.D. diss., Cambridge Univ.

\bibitem[Stevens et al.(1992)]{SB92} 
Stevens, I.~R., Blondin, J.~M., \& Pollock, A.~M.~T.\ 1992, \apj, 386, 265 

\bibitem[Struve(1950)]{Struve50}
Struve, O.\ 1950, {\it Stellar Evolution}, Princeton University Press, NJ. 

\bibitem[Tassoul(1987)]{Tassoul87} 
Tassoul, J.-L.\ 1987, \apj, 322, 856 

\bibitem[Thomas et al.(2005a)]{Thomas05a}
Thomas, B.~C., et al.\ 2005a, \apj, 634, 509

\bibitem[Thomas et al.(2005b)]{Thomas05b} Thomas, B.~C., Jackman,
C.~H., Melott, A.~L., Laird, C.~M., Stolarski, R.~S., Gehrels, N.,
Cannizzo, J.~K., \& Hogan, D.~P.\ 2005b, \apjl, 622, L153

\bibitem[Torres et al.(1986)]{TCM86} 
Torres, A.~V., Conti, P.~S., \& Massey, P.\ 1986, \apj, 300, 379 

\bibitem[Tuthill et al.(1999)]{nature99} 
Tuthill, P.~G., Monnier, J.~D., \& Danchi, W.~C.\ 
1999, \nat, 398, 487 

\bibitem[Tuthill et al.(2000)]{keckmask} 
Tuthill, P. G., Monnier, J. D., Danchi, W. C., Wishnow, E. H., \& Haniff, C. A. 
2000, \pasp, 112, 555

\bibitem[Tuthill et al.(2000)]{Tuthill00} 
Tuthill, P.~G., Danchi, W.~C., Hale, D.~S., Monnier, J.~D., \& Townes, C.~H.\ 
2000, \apj, 534, 907 

\bibitem[Tuthill et al.(2002)]{Tuthill02} 
Tuthill, P.~G., Monnier, J.~D., Danchi, W.~C., Hale, D.~D.~S., \& Townes, C.~H.\ 
2002, \apj, 577, 826 

\bibitem[Tuthill et al.(2002)]{wrconf02} 
Tuthill, P.~G., Monnier, J.~D., \& Danchi, W.~C.\ 2002, 
Interacting Winds from Massive Stars, 260, 321 

\bibitem[Tuthill et al.(2002)]{redrect02} 
Tuthill, P.~G., Men'shchikov, A.~B., Schertl, D., Monnier, J.~D., Danchi, W.~C., \& 
Weigelt, G.\ 2002, \aap, 389, 889 

\bibitem[Tuthill et al.(2003)]{wrconf03} 
Tuthill, P.~G., Monnier, J.~D., Danchi, W.~C., \& Turner, N.~H.\ 
2003, A Massive Star Odyssey: From Main Sequence to Supernova, 212, 121 

\bibitem[Tuthill et al.(2006)]{Quint06} 
Tuthill, P., Monnier, J., Tanner, A., Figer, D., Ghez, A., \& Danchi, W.\ 2006, 
Science, 313, 935 

\bibitem[Usov(1991)]{Usov91} 
Usov, V.~V.\ 1991, \mnras, 252, 49
 
\bibitem[Walder \& Folini(2000)]{WF00} 
Walder, R., \& Folini, D.\ 2000, \apss, 274, 343 

\bibitem[Woosley et al.(2002)]{WHW02} 
Woosley, S.~E., Heger, A., \& Weaver, T.~A.\ 2002, Reviews of Modern Physics, 74, 1015 

\bibitem[Woosley \& Bloom(2006)]{WB06} 
Woosley, S.~E., \&  Bloom, J.~S.\ 2006, \araa, 44, 507 

\bibitem[Vanbeveren et al.(1998)]{VLV98} 
Vanbeveren, D., De Loore, C., \& Van Rensbergen, W.\ 1998, \aapr, 9, 63 

\bibitem[Williams et al.(1990)]{Williams90} 
Williams, P.~M., van der Hucht, K.~A., Pollock, A.~M.~T., Florkowski, D.~R., 
van der Woerd, H., \& Wamsteker, W.~M.\ 1990, \mnras, 243, 662 

\bibitem[Williams(1997)]{Williams97} 
Williams, P.~M.\ 1997, \apss, 251, 321 

\bibitem[Williams \& van der Hucht(2000)]{WV00} 
Williams, P.~M., \& van der Hucht, K.~A.\ 2000, \mnras, 314, 23 

\bibitem[Zahn(1977)]{Zahn77} Zahn, J.-P.\ 1977, \aap, 57, 383

\bibitem[Zubko(1998)]{Zubko98} 
Zubko, V.~G.\ 1998, \mnras, 295, 109 

\end{thebibliography}
\end{document}